\documentstyle[twocolumn,psfig,aps]{revtex}
\begin{document}
\draft
\twocolumn[\hsize\textwidth\columnwidth\hsize\csname @twocolumnfalse\endcsname
%
%
%

\title{Qualitative understanding of the sign of t$'$ asymmetry\\
 in the extended t-J Model and relevance for pairing properties}

\author{G. B. Martins$^1$, J. C. Xavier$^2$, L. Arrachea$^3$ and E. Dagotto$^1$}

\address{$^1$ National High Magnetic Field Lab and Department of Physics,
Florida State University, Tallahassee, FL 32306, USA}

\address{$^2$ Instituto de Fisica Gleb Wataghin, Universidade Estadual de Campinas,
Campinas, SP, Brasil}

\address{$^3$ Departamento de Fisica, Universidad de Buenos Aires, FCEN, Buenos Aires, DF, Argentina}

\date{\today}
\maketitle

\begin{abstract}
Numerical calculations illustrate the effect of the sign of the next nearest-neighbor 
hopping  term t$'$ on the 2-hole properties of the t-t$'$-J model. Working mainly on 
2-leg ladders, in the -1.0 $\leq$t$'$/t $\leq$ 1.0 regime, it is shown that introducing 
t$'$ in the t-J model is equivalent to effectively renormalizing J, namely 
t$'$ negative (positive) is equivalent to an effective t-J model with smaller (bigger) J. 
This effect is present even at the level of a 2$\times$2 plaquette toy model, 
and was observed also in calculations on small square clusters. Analyzing the transition 
probabilities of a hole-pair in the plaquette toy model, it is argued that the coherent 
propagation of such hole-pair is enhanced by a constructive interference between both t and t$'$ for 
t$'>$0. This interference is destructive for t$'<$0.
\end{abstract}
\vskip2pc]
\narrowtext

%
%

\medskip

One of the most important unsolved problems in theoretical physics is the 
clarification of the nature of high temperature superconductors. 
A popular approach in this context is the use of the t-J model, with holes moving 
in an antiferromagnetic (AF) spin background. In recent years, mainly due to 
an increase in the sensitivity and resolution of 
angle resolved photoemission spectroscopy (ARPES), it has been shown that 
extra hole hoppings beyond nearest-neighbor (NN) are important in the t-J model, 
giving origin to the ``extended'' t-J model. 
For example, ARPES measurements in Sr$_{2}$CuO$_{2}$Cl$_{2}$\cite{wells}, and 
their subsequent interpretation\cite{nazarenko}, have shown the importance of 
those extra hoppings to reproduce the experimental results. Subsequent efforts have 
concentrated on the effect of the extra hoppings on various properties 
of planar and ladder systems, such as stripe stability\cite{tohyama1}, competition between 
pairing and stripes\cite{white1}, spin-charge separation in 2-d\cite{martins1}, 
stripe formation mechanism\cite{martins2}, spin gap evolution\cite{poilblanc}, and 
current-current correlations\cite{tsutsui}. Most of these papers 
have compared and contrasted the dependence of different properties of the 
extended t-J model with the sign of the next NN (NNN) hopping t$'$. Currently it is well established 
that a positive t$'$ enhances hole pairing and AF correlations, 
while the opposite occurs for t$'$ negative\cite{white1,tohyama2}. Nevertheless, to the best of our 
knowledge, these previous publications have not provided an intuitive 
mechanism that can explain why this happens, namely, for what reason there is an asymmetry 
between positive and negative t$'$. This is particularly puzzling considering the limit t=0, 
since in the t$'$-J model the sign of t$'$ is irrelevant\cite{independent}. 

It is the purpose of this paper 
to provide a qualitative explanation to this phenomenon, i. e., the sign of t$'$ asymmetry. 
Our main result is that a quantum interference 
between NN and NNN hoppings identified in the hole-pair propagation was found to be constructive 
(destructive) for t$'$ positive (negative); this accounts for the observed dependence of the 
hole-pair properties with the sign of t$'$. The t-t$'$-J model used here is defined as
$$
\rm
H = J \sum_{\langle {\bf ij} \rangle} 
({{{\bf S}_{\bf i}}\cdot{{\bf S}_{\bf j}}}-{{1}\over{4}}n_{\bf i}n_{\bf j})
- \sum_{ {\bf im} } t_{\bf im} (c^\dagger_{\bf i} c_{\bf m} + h.c.),
\eqno(1)
$$
where $\rm t_{\bf im}$ is t for NN, t$'$ for NNN, and
zero otherwise. The rest of the notation is standard.
Comparison with ARPES experiments\cite{nazarenko,arpes} showed that t$'$$<$0 is 
physically relevant for the hole-doped cuprates.
The Density Matrix Renormalization Group (DMRG)\cite{white2} 
and Lanczos\cite{review} methods are used on ladders and small square clusters to study the 
Hamiltonian Eq. (1).

First, let us show that the dependence of hole-hole correlations with the sign of 
t$'$ can be crudely described as a renormalization of the exchange interaction J.
In Fig. 1, it is shown, through calculations on ladders and square clusters, 
the dependence of the average distance $\langle$d$\rangle$ between two holes with the sign of t$'$. 
The result obtained is roughly consistent with a renormalization of J by t$'$, in the 
sense that results for t$'$ negative (positive) can be obtained by renormalizing J to a 
smaller (bigger) effective value, leading to an increase (decrease) in $\langle$d$\rangle$. 
To show that binding energy and phase separation (PS) tendencies are both affected in a way consistent 
with this interpretation it is shown in Fig. 1.c (squares) the dependence of 
the PS line with t$'$. It can be observed that t$'$ negative (positive)
requires an increase (decrease) in the value of J needed 
for the holes to segregate (desegregate), if compared to the 
J value that leads to PS at t$'$=0. The circles 
display values of J/t and t$'$/t that result in the holes having a binding energy of $\approx$ 0.5t 
(as the boundary of the binding region that we can consider ``robust''). This 
binding energy line shows that at a fixed J, such as 0.4, increasing t$'>$0 leads to strong binding, with the 
opposite effect for t$'<$0. As expected, this line approximately follows the behavior of the PS 
line\cite{ps}. Thus, the essence of the results in previous studies\cite{white1,tohyama2}, showing pairing with 
t$'>$0, can be reproduced on a small cluster with 2 holes. 
Note that in Fig. 1, for t$'$/t$\approx$1, $\langle$d$\rangle$ reaches its minimum 
value and starts to increase. In the limit when $|$t$'|$/t$\gg$1 the hole-hole correlations 
(and other properties of the model) become {\it independent} of the sign of t$'$\cite{independent}. 
Therefore, the renormalization 
of J described above mainly occurs in the region -1.0$\lesssim$t$'$/t$\lesssim$1.0.

\begin{figure}
\begin {center}
\mbox{\psfig{figure=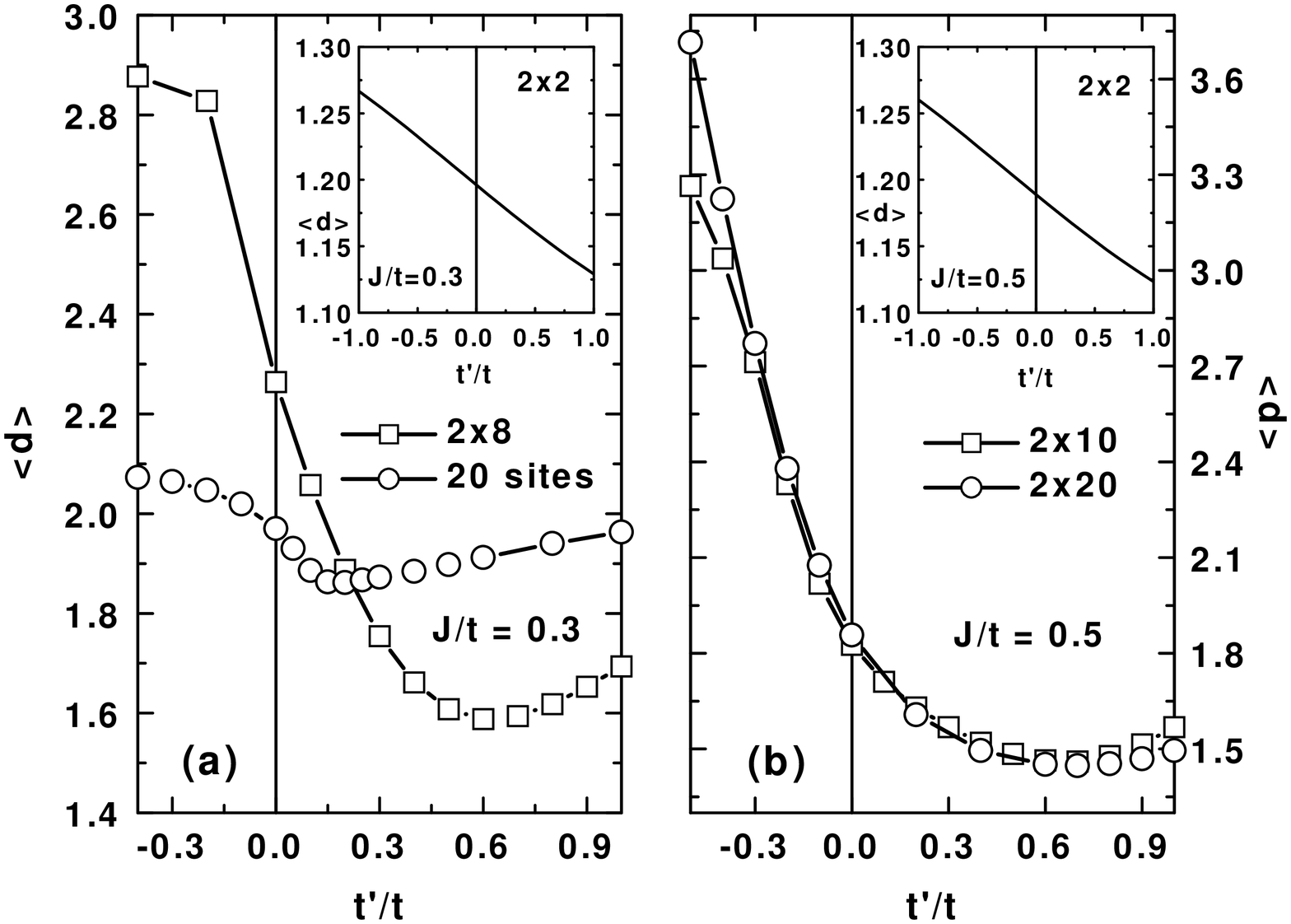,width=3.0in}}
\mbox{\psfig{figure=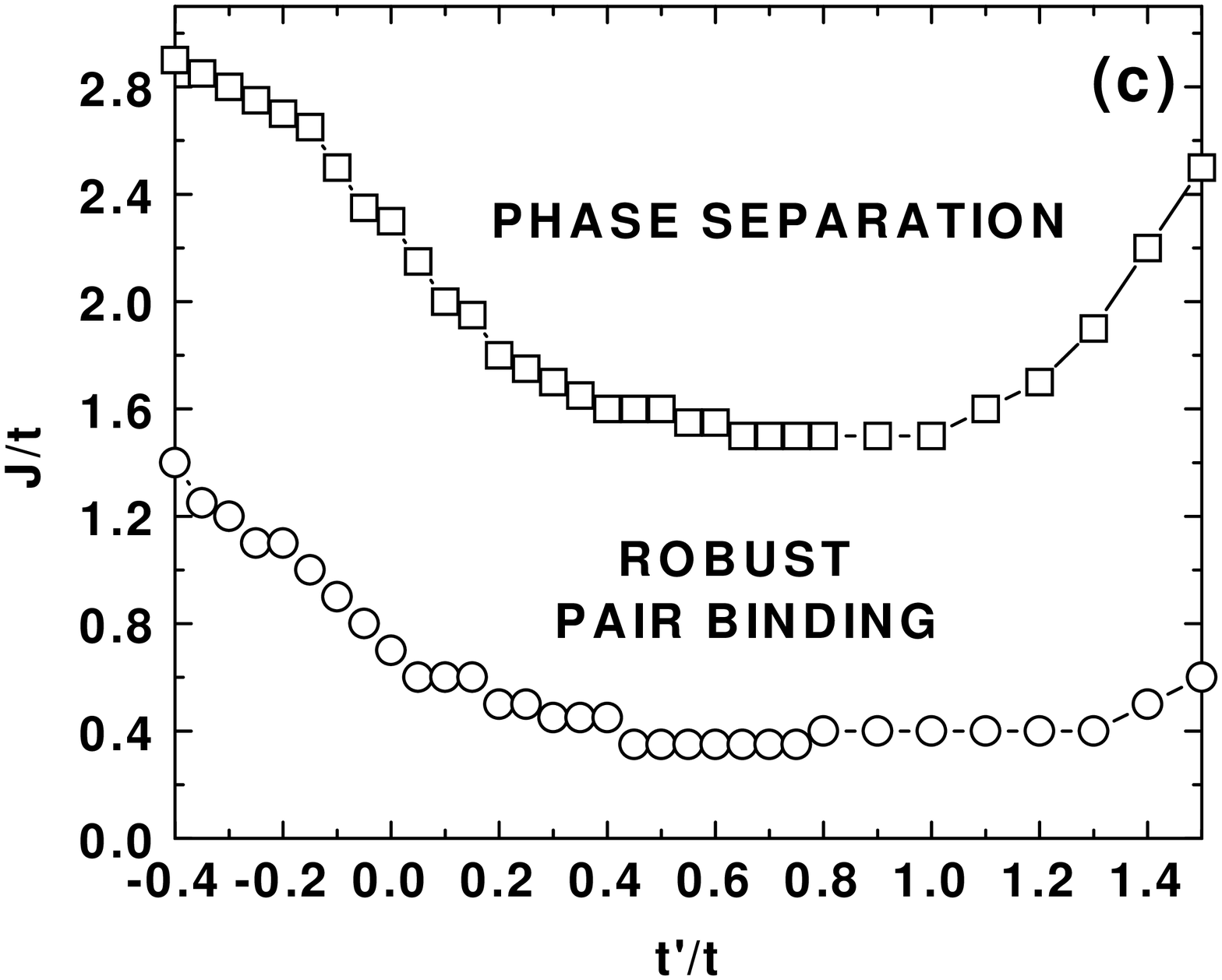,width=2.2in}}
\end{center}
\caption{(a) Exact diagonalization (ED) results showing the dependence with t$'$ of the average distance $\langle$d$\rangle$ 
between two holes on a 2$\times$8 
ladder (squares) and on a 20 sites tilted 2-d cluster (circles). Periodic boundary conditions (PBC) are used in both cases, 
J/t=0.3 and -0.5$\leq$t$'$/t$\leq$1.0. The tendency of the holes to separate 
when a negative t$'$ is turned on can be observed. For t$'>$0, there is a tendency of the holes to form 
tighter pairs. However, in this last case, as t$'$ keeps on increasing the holes will eventually 
tend to separate, showing a similar behavior to the t$'$ negative 
case. The inset shows a calculation of $\langle$d$\rangle$ on a 2$\times$2 plaquette with conclusions similar 
to those reached with the larger clusters. (b) Same as (a) except that 
now J/t=0.5 on a 2$\times$10 ladder (ED) and the tilted cluster was substituted by a 2$\times$20 ladder 
(DMRG) with open boundary conditions (OBC). Again the inset shows results of $\langle$d$\rangle$ on 
2$\times$2, but now for J/t=0.5. (c) Phase diagram J/t 
vs. t$'$/t showing regions of pair binding and phase separation (defined through the divergence of the compressibility). 
The pair binding line (circles) is defined 
by values of J/t and t$'$/t that give a robust binding energy of $\approx$ 0.5t on a 2$\times$8 ladder. Notice 
that close to t$'$/t=0 both lines behave in accordance with our qualitative picture, i. e., t$'$ negative (positive) 
renormalizes J to smaller (bigger) values.
} 
\end{figure}

It is important to note that it is not only $\langle$d$\rangle$ that behaves in accordance 
with this simple scenario. To a surprising accuracy, hole-hole correlations for, e. g., 
t$'$ negative in the t-t$'$-J model, match those of the t-J model with an 
effective (smaller) J. To illustrate that, on Fig. 2 hole-hole correlations for a 2$\times$8 
ladder with 2 holes are calculated for the t-t$'$-J model with J/t=0.2 and t$'$/t=-0.2, 
and then compared to results for the t-J model with J/t=0.07. The open circle stands for 
a projected hole (at the origin of the coordinate system) and the radius of the solid 
circle on site {\it i} is proportional to $\langle$n$_{0}$n$_{i}$$\rangle$. Most of the 
values match to high accuracy, as deduced from the similarity of the pictures. This 
is also the case for t$'$ positive and for square clusters. Then, this seems an indication 
that the renormalization concept is robust. Preliminary results indicate that 
the main features of the dependence of the spin correlations with the sign of t$'$ 
can be qualitatively explained through J's renormalization, although it is not possible to 
achieve the same degree of high accuracy as shown above in the charge sector.

\begin{figure}
\begin {center}
\mbox{\psfig{figure=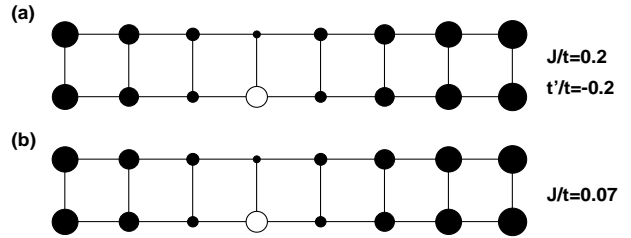,width=3.2in}}
\end{center}
\caption{Comparison of hole-hole correlations $\langle$n$_{0}$n$_{i}$$\rangle$ between the 
t-t$'$-J model and an effective 
t-J model. (a) 2$\times$8 ladder with 2 holes, J/t = 0.2 and t$'$/t = -0.2, PBC. Open circle 
stands for a projected hole at origin, while solid circles at site $i$ have radius proportional to 
$\langle$n$_{0}$n$_{i}$$\rangle$. (b) Same as (a) but for J/t = 0.07 and t$'$/t = 0.0.
} 
\end{figure}

The fact that the renormalization of the exchange interaction by t$'$ is 
consistently observed on ladders and square clusters (compare result for 2x8  ladder and 
20 sites cluster on Fig. 1.a), 
and is observed for OBC and PBC 
(compare result for 2x10 PBC ladder with 
2x20 OBC ladder on Fig. 1.b), is an indication that this effect  
is associated with some local process and therefore could even be observable on 
a 2$\times$2 plaquette with 2 holes. If this is correct then it should be possible, 
through a careful analysis of such toy model, to gain a better insight 
on the qualitative aspects of the physics associated with the sign of 
t$'$. In the insets for Fig. 1 it is shown the dependence of $\langle$d$\rangle$ 
with t$'$ on a 2$\times$2 plaquette and the same trend found on ladders and square clusters is again 
reproduced. Also, the exact energy of the ground state for the plaquette 
with two holes is given by $E_{0}=-\frac{1}{2}\left[ \left( J+2t^{\prime }\right) +%
\sqrt{\left( J+2t^{\prime }\right) ^{2}+32t}\right] $, where the renormalization of 
J by t$'$ can be explicitly seen. It is important to remark that in a reduced basis formed with spin singlets 
on the sides and diagonals of the plaquette, the coefficients 
describing the ground state also depended on t$'$ through the same 
expression $\left( J+2t^{\prime }\right)$.

Encouraged by these results on the 2$\times$2 plaquette, one can now  
try to understand qualitatively how the change of sign in one hopping amplitude 
can change the binding properties of a hole-pair, a result that up to 
now is being rephrased as a renormalization of J. To make progress it has to 
be analyzed how this change of sign affects the dynamics of a hole-pair. 
A hint in this direction is that for t=0 this asymmetry in the 
properties of the model caused by the sign of t$'$ vanishes, as implied 
above by the discussion of the $|$t$'|$/t$\gg$1 regime. Then, for the hole-pair properties 
to be affected by the sign of t$'$ it is important to have the possibility of 
NN t hoppings. Intuitively this resembles an interference of some sort: the movement 
of a hole-pair, through a combination of both hoppings, may lead to a 
coherent propagation of the pair, for t$'$ positive, or to its melting into independent 
quasi-particles\cite{poilblanc,troyer}, for t$'$ 
negative\cite{spincharge}. To check this idea, in the plaquette toy model, one can calculate the probability 
of a transition from an initial state composed of a hole-pair and a spin singlet in 
opposite sides of the plaquette, to a final state where the hole-pair and the spin 
singlet have exchanged places\cite{strong}. Such a 
transition is depicted in Fig. 3, where in (a) it is shown the most probable 
second order process (a process with two t hoppings would be less probable because 
the intermediate state would be an excited state\cite{strong}) and in (b) it is shown a third 
order process (the other three are equivalent to the one discussed). 
The difference between processes (a) and (b) resides in the fact that the latter 
needs one more virtual state than the former. As such state is excited ($\Delta$E=J), 
and taking in account that there are three other third order processes 
with two NN hoppings and one NNN 
hopping, it can be shown that second and third order processes will have 
amplitudes proportional to t$'^{2}$ and 4t$^{2}t'/$J, respectively. This means that if t$'$ is positive 
(negative) they will have the same ($\pi$-shifted) 
phase and their interference will be constructive (destructive). A similar reasoning can be 
applied to higher order processes, but it can be 
shown that they are less probable than the processes discussed above, given that 
they would pass by the same virtual state more than once\cite{tohyama3}.

\begin{figure}
\begin {center}
\mbox{\psfig{figure=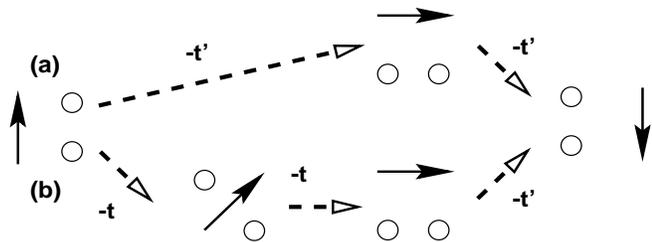,height=1.2in,width=3.3in}}
\end{center}
\caption{(a) Second order process that exchanges the positions of a hole pair (circles) and a spin singlet 
(solid arrow) localized on opposite sides of a 2$\times$2 plaquette. (b) Third order process for the 
same transition depicted in (a). Both processes will have the same amplitude 
if t$'=$ 4t$^{2}$/J$>$0 (see text). This leads to a constructive interference if t$'$ is positive, which becomes destructive 
if t$'$ is negative.
} 
\end{figure}

Through the mechanism described in Fig. 3, it is suggested that  achieving coherency 
in the propagation of the hole-pair depends on having the 
correct balance of short-range processes (1 and $\sqrt{2}$ hoppings). 
Then, it should not be a surprise that the plaquette can display this 
effect, as shown above. Nevertheless, it should be checked that such process also occurs on 2$\times$L ladders. 
That this is the case is shown on Fig. 4, 
where pair-field correlations at distance of one lattice spacing are calculated through 
ED on 2x10 ladders with J/t = 0.5 and $\langle$n$\rangle$ = 0.9. The pair operator is defined as 
$\Delta_{i} =c_{i1\uparrow }c_{i2\downarrow }-c_{i1\downarrow }c_{i2\uparrow }$, 
where $i$ labels a rung and the legs are numbered 1 and 2. The result 
shows that the coherent propagation of a hole-pair located in a rung is 
enhanced for t$'$ positive, while a rapid decay 
is observed for t$'$ negative, in agreement with the picture described in 
the toy model and with previous calculations\cite{white1,tohyama2}.

\begin{figure}
\begin {center}
\mbox{\psfig{figure=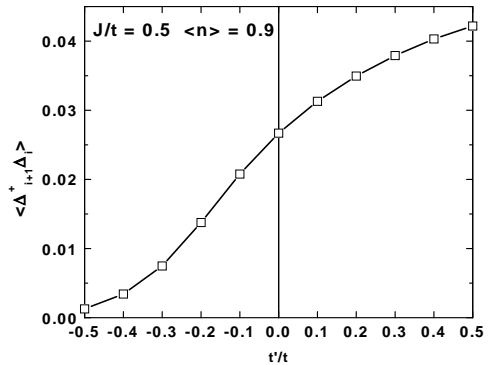,width=3.in}}
\end{center}
\caption{Dependence with t$'$/t of the pair-pair correlations at distance 1 in a 2$\times$10 
ladder with 2 holes for J/t = 0.5. The coherent propagation of a hole-pair located in 
a rung is shown to have a dependence with t$'$/t consistent with the scenario described 
on Fig. 3.
} 
\end{figure}

The physics of the t$'>$0 extended t-J model resembles results for
the effective model discussed in Ref.\cite{afvh}, where
holes were considered quasi-particles moving in a ``perfect''
antiferromagnetic background with hopping only within the
same sublattice, and with an explicit NN attraction to mimic
AF mediated pairing. In fact, in Ref.\cite{naza} it was shown
that a {\it positive} t$'$ is needed to generate a $d_{x^2-y^2}$
two-hole bound state in the quasi-particle model 
of Ref.\cite{afvh}. As a consequence, 
the regime of t$'>$0 of the extended t-J model,
with its strong AF correlations and pair coherent movement,
is likely mimicked by the simple toy model used in previous
literature\cite{afvh,naza}.

Summarizing, here it has been provided a qualitative picture to explain the 
dependence of hole-pairing on the sign of the NNN hoping t$'$ in the t-t$'$-J model. 
Through numerical calculations on square clusters, but mainly on ladders, using ED and 
DMRG, it was established that t$'$ negative (positive) effectively reduces (increases) J. 
The variety of clusters and boundary conditions where this effect was consistently 
observed served as an indication of the locality of the process involved. This suggested the use of 
a 2$\times$2 cluster with 2 holes as a guiding toy model. By solving analytically 
this cluster, the J renormalization was shown explicitly in the dependence of the ground 
state energy with t$'$, and the behavior of hope-pair size $\langle$d$\rangle$ was consistent with 
what was found on ladders and square clusters. The fact that for t=0 the properties of the 
t-t$'$-J model are not dependent on the sign of t$'$ has indicated that some sort of interference 
process between t and t$'$ should be responsible for the hole-pair dependence on t$'$. By analyzing 
transition probability amplitudes in the plaquette it was observed that indeed this is 
the case. A negative t$'$ interferes destructively with t, causing a hole-pair to have a tendency 
to split into two independent quasi-particles. A positive t$'$, on the other hand, by interfering 
constructively with t, preserves the integrity of the hole-pair while it propagates. This simple 
picture provides a better understanding of the t-t$'$-J model, adding more physical 
insight to the ``effective J'' picture\cite{martins1}. By calculating pair-field correlations at distance one 
on 2$\times$10 ladders it was argued that this qualitative explanation holds also for larger systems.

The authors wish to thank NSF (DMR-9814350) and FAPESP-Brazil for finacial support. 
GBM and JCX acknowledge fruitful discussions 
with A. Feiguin, C. Gazza, A. Malvezzi and J. Riera.




\begin{references}
\bibitem{independent} 
It can be shown that changing the sign of t in the t-J model is equivalent to a shift in 
momentum by $\left( \pi ,\pi \right) $. Thus, by the expression of the dispersion relation 
of a hole in the t-J model $0.55J\left(\cos k_{x}+\cos k_{y}\right)^{2}$ it can be concluded 
that such a change of sign has no effect in the properties of the model. Changing the sign of 
t$'$ in the t$'$-J model is also irrelevant; this can be concluded by noticing that the t$'$-J model 
is equivalent to a t-J model on a $\sqrt{2}$a lattice. Therefore, in the t-t$'$-J model, in the  
limit where either t=0 or t$'$=0, a change of sign in the hopping amplitude is irrelevant.
\bibitem{spincharge}
In previous works by two of the authors (ref. \cite{martins1}) indications of spin-charge 
separation were found in the negative t$'$ regime. The connection of the ideas discussed here 
to these previous results is being presently analyzed.
\bibitem{strong}
Since the objective is to analyze the hole-pair dynamics, a strong coupling regime 
(J $\gg$ t, t$'$) is assumed. In this case the two states 
with spin singlets in the diagonal of the plaquette are excited configurations (with zero energy) and the four states 
with spin singlets in the sides of the plaquette form a degenerate ground state (with 
energy -J). Therefore, starting from a ground state configuration (spin singlet on the side), a NN hopping takes it 
to an excited configuration (spin singlet on the diagonal), being therefore a less probable process, 
according to standard perturbation theory. Note that the fermionic commutation signs are included 
in the orientation of the singlets.
\end{references}
\end{document}